\documentclass[slac_one]{revtex4}
\usepackage{graphicx}
\usepackage{color}
\usepackage{fancyhdr}
\begin{document}

\title{Searching for New Physics: Results from Belle and Babar
\footnote{University of Cincinnati preprint \# UCHEP-07-01}
\footnote{Talk presented at XXXIV SLAC Summer Institute, Stanford Linear Accelerator Center, July 17-28, 2006}} 

%

\author{K. Kinoshita}
\affiliation{University of Cincinnati, Cincinnati, OH 45221, USA}

\begin{abstract}
The $B$-factories provide rich opportunities to search for new phenomena, in $B$, charm, and tau decays.  
Presented here is a selection of recent results from Belle and Babar.
\end{abstract}

\maketitle

\thispagestyle{fancy}


\section{SEARCH STRATEGIES FOR NEW PHYSICS}
The Standard Model (SM) of fundamental particles and forces describes beautifully nearly all phenomena and rates that have been measured, in experiments at accelerators and elsewhere.
What it does {\it not} do is $(i)$ incorporate the gravitational force and $(ii)$ explain a few crucial fundamental questions, such as the baryon asymmetry of the universe and the origin of mass.
Without these, our understanding is not complete, and because of this we continue to search for a more general theory, ``Beyond the Standard Model'' (BSM), that will encompass the SM and additionally contain ``New Physics'' (NP) elements that address the remaining questions.

In general, NP involves new particles that couple to known particles in ways that would not have been detected previously.  
Due to constraints from existing data, most candidate "New Physics" particles tend to occur in the mass range above $\sim$100~GeV or couple very weakly to normal matter.
One approach to search, which could be called ``brute force,'' is simply to look in the highest energy collisions.
More subtly, new particles can cause interaction rates to deviate from expectations by appearing in virtual intermediate states at energies below their mass.
In terms of search strategies, it makes sense to concentrate on areas where any ``New Physics'' (NP) is most likely to be recognized, that is, where the SM value is finite and can be precisely measured or where it is highly suppressed or forbidden.

At the $B$-factory there are many such opportunities to observe possible deviations from the SM.
In $B$-decays, complex phases and magnitudes of CKM matrix elements are measured, many of them by multiple approaches with varying sensitivities to NP.
$B$-factories also produce large samples of charm, where strong suppression of mixing and $CP$-asymmetries in the SM provide windows to NP.
Of comparable magnitude is the tau pair sample, which provides opportunities to search for phenomena that violate lepton flavor and/or baryon number.

\section{THE B-FACTORY EXPERIMENTS}
The results presented here are based on data collected by the Belle\cite{belle} the Babar\cite{babar} detectors at the KEKB\cite{kekb} and PEP2\cite{pep2} asymmetric $e^+e^-$ storage rings, respectively.
As of the week of this presentation (July 24, 2006), Belle and Babar had collected 630~fb$^{-1}$ and 371~fb$^{-1}$ of data ($\sim 90\%$ at the $\Upsilon$(4S) resonance, $\sim$10\% at a lower, nonresonance energy), for a combined sample of over 10$^9$ $B\bar B$ events.
In addition to $B\bar B$ events, these samples contain a comparable number of continuum charm ($e^+e^-\to c\bar c$) and tau pair events.
Belle has also collected data at the $\Upsilon$(5S) resonance ($\sqrt{s}$=10.869~GeV), and one result from 1.86~fb$^{-1}$ of data, containing $9\times 10^4$ $B_s^{(*)}\bar B_s^{(*)}$ events,  will be reported here. 

\section{THE ``OLD PHYSICS''}
In order to recognize ``New Physics,'' it is first useful to understand the ``Old Physics.''
Here I give a very brief description of aspects of CKM physics that are crucial to understanding the significance of the results to be described.

The weak charged-current couplings of the quarks, arranged in the $3\times 3$ matrix known as the CKM matrix,
describe in the SM the transformation between the mass and weak eigenstates of the quarks.
As a transformation between two complete sets of eigenstates, the matrix must be complex as well as preserving the orthogonality and metric, {\it i.e.}, it must be unitary.   
Formally, the elements $\{V_{ij}\}$ of a unitary matrix must satisfy
$\sum_{j}V_{ji}^*V_{jk}=\delta_{ik}$,
and for a $3\times 3$ matrix the constraints imposed by these conditions reduce the freedom of CKM to four parameters, three real and one irreducibly complex, often represented explicitly as
\begin{eqnarray}
{\cal M}=
\left(\begin{array}{ccc}
V_{ud}&V_{us}&V_{ub}\\
V_{cd}&V_{cs}&V_{cb}\\
V_{td}&V_{ts}&V_{tb}
\end{array}\right)
=
\left(\begin{array}{ccc}
1-\lambda^2/2& \lambda  & \lambda^3A(\rho -i\eta)\\
-\lambda & 1-\lambda^2/2 & \lambda^2 A\\
\lambda^3A(1-\rho -i\eta) & -\lambda^2 A & 1
\end{array}\right)
\label{eqn:ckm}
\end{eqnarray}
The unitarity condition, applied to $\{i=1,k=3\}$, results in
\begin{eqnarray}
0&=&{V_{ub}^*V_{ud}\over V_{cb}^*V_{cd}}+1+ {V_{tb}^*V_{td}\over V_{cb}^*V_{cd}}\label{eq:UT}\\
&\approx& -(\rho+i\eta)+1-(1-\rho-i\eta).\nonumber
\end{eqnarray}
This sum of three terms may be represented as a closed triangle in the complex plane with corners at $(0,0)$, $(1,0)$ and $(\rho,\eta)$.
In the context of the $B$-factories, this has come to be known as ``{\it The} Unitarity Triangle'' and embodies the least precisely known aspects of the CKM matrix.
The principal objective of the $B$-factory experiments is to test the validity of Equation~(\ref{eq:UT}), {\it i.e.} the closure, or self-consistency, of this triangle.

\subsection{CP asymmetry in B decay: example}
A process with a complex coupling constant is intrinsically $CP$-violating.
In the CKM paradigm the irreducible complexity of the matrix occurs only with three or more generations, so it stands to reason that all three generations must be involved in any process where $CP$-asymmetry is to be observed.  
Because $b$-hadrons include a third-generation quark, their decays often satisfy this criterion.
However, as the transition rate is proportional to the absolute square of the amplitude, any decay that proceeds by a single process can not exhibit $CP$-asymmetry.
For the $CP$-asymmetry to be physically observable, a decay or interaction must proceed by at least two separate processes; if their amplitudes are $g{\cal A}$ and $g'{\cal A}'$ (where the CKM couplings are factored out as $g$ and $g'$), then the rate is proportional to $|g{\cal A}+g'{\cal A}'|^2$, whereas the rate for the $CP$-inverted state is $|g{\cal A}^*+g'{\cal A}'^*|^2$.
These are unequal, i.e., $CP$ symmetry is violated, if $g$ and $g'$ have non-zero relative complex phase.

\begin{figure}[t]
\centering
\includegraphics[height=25mm]{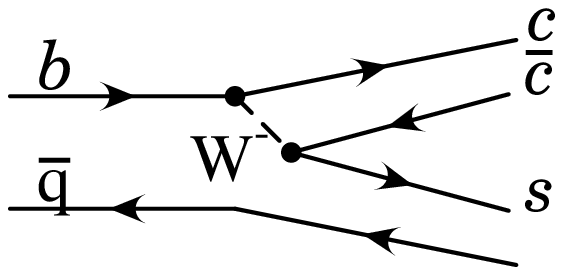}
\includegraphics[height=25mm]{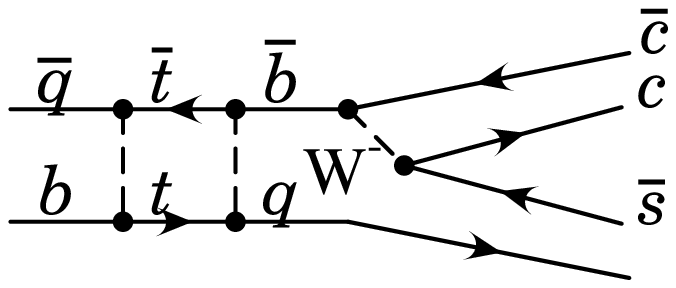}
\caption{Processes for decay $B\to J/\psi K_S$, (left) tree and (right) mixing + tree.} 
\label{fig:jpsiks}
\end{figure}
The ``golden'' example for $CP$-asymmetry in $B$ decay is in the mode $B\to J/\psi K_S$, which has $CP=-1$.
The two interfering processes are shown in Figure~\ref{fig:jpsiks}.
For the ``tree'' process (left) $\bar B^0\to J/\psi K_S$, the amplitude is proportional to the product $V_{cb}^*V_{cs}$, which in the parameterization of eq.(\ref{eqn:ckm}) is real.
The decay may also proceed through a mixing oscillation, $\bar B^0\to B^0\to J/\psi K_S$ (Figure~\ref{fig:jpsiks}(right)), for which the amplitude is proportional to $V_{tb}^{*2}V_{td}^2 V_{cb}V_{cs}^*$.
As the ``tree'' parts of both processes are identical (except for a particle conjugation), the hadronic components of the two amplitudes are the same.  
The two amplitudes differ in the mixing, which is well-measured and understood, and in their complex phase by the phase of $V_{tb}^{*2}V_{td}^2$, which is expressed in terms of the angles of Unitarity Triangle as $2\phi_1$ or $2\beta$.
The result is a $CP$-dependent decay rate
\begin{eqnarray}
{dN\over dt}(B\rightarrow f_{CP})={1\over 2}\Gamma e^{-\Gamma\Delta t}
[1+\eta_b\eta_{CP}{\rm sin}2\phi_1{\rm sin}(\Delta m\Delta t)],
\label{eqn:cpasym}
\end{eqnarray}
where $\eta_b=+1(-1)$ for a $B^0(\bar B^0)$,
$\eta_{CP}=+1(-1)$ if CP is even (odd), and $\Delta t$ is the time 
interval from creation to the CP eigenstate decay.

Here I present a very brief outline of the analysis procedure for time-dependent CP anaylsis, which is common to many of the results presented here.
A sample of $CP$-eigenstate decays is first reconstructed, e.g., $\bar B^0\to J/\psi K_S$ (inclusion of charge conjugate decays is implied throughout this article).
The signal is observed through the distributions of candidates in $\Delta E$ and $M_B$; 
 $\Delta E$, the difference between the candidate's reconstructed energy (in the collision center-of-mass (CM)) and the CM beam
energy, is centered at zero for signal with a width that is dominated by detector resolution
(10-50 MeV), and $M_B=\sqrt{E_{beam}^{*2}-p^{*2}_{cand}}$, known as the beam-constrained
mass, has a width determined by detector momentum resolution and accelerator beam energy
spread.  
For each event containing a good signal candidate, known as a $CP$ tag, the remainder of the event is 
examined to determine the original flavor ($b$ or $\bar b$) of the other $B$, using what is known as ``flavor tagging'';
the flavor is correlated with the charge sign of tracks from $b$ decay.  
Useful correlations include: high-momentum ($p^*>1.1$~GeV/c) electrons 
or muons from $b\rightarrow c\ell^-\bar\nu$, $K^-$ from $b\rightarrow 
cX \{c\rightarrow sY$\}, lower momentum leptons from $b\rightarrow cX 
\{c\rightarrow \ell^+Y$\}, and soft pions from $b\rightarrow D^{*+}X 
\{D^{*+}\rightarrow D^0\pi^+$\}.
To determine the decay time difference, $\Delta t$, we measure the decay vertices of both $B$ decays
in each tagged event.
The vertex of the flavor-tagged $B$ is estimated by examining the tracks not associated
with the CP decay (excluding off-axis tracks such as those from identified $K_S$).
The resolution of the vertex on the flavor-tag side is limited not only by detector resolution but also by
the event itself, which may contain secondary charm particles and thus not have a unique
vertex.
$\Delta t$ is calculated as the measured 
difference in the reconstructed z-coordinates, $\Delta z$, divided by 
$\beta\gamma c$ ($\beta\gamma=0.425$ at Belle, 0.488 at Babar) the Lorentz boost of the beam collision center-of-mass.

An unbinned maximum likelihood fit to sin2$\phi_1$ is then performed 
on the distribution in $\Delta t$,
accounting for the CP and flavor of each event.
To make this measurement one must account for the backgrounds to the 
CP tag, the fraction of incorrect flavor tags, and various sources of 
uncertainty on the measurement of $\Delta t$.
The fitting function takes into account the root distribution of the 
signal (an analytic function, eq.~(\ref{eqn:cpasym})), the fraction of incorrect flavor tags, background to reconstructed CP decays (both 
correctly and incorrectly tagged), and the resolution of $\Delta t$, 
parametrized and assigned event-by-event.

\section{MEASUREMENTS WITH SENSITIVITY TO NEW PHYSICS }

Due to the unfortunate timing of this Institute vis-a-vis the ICHEP06 meeting this year, very few new results are able to be shown here.  
Most of these results are thus based on results as of the Winter 2006 conferences, and updated results will be released next week.

\subsection{$CP$ Asymmetry in $\bar b\to ss\bar s$}

\begin{figure}[tbp]
\centering
\includegraphics[height=30mm]{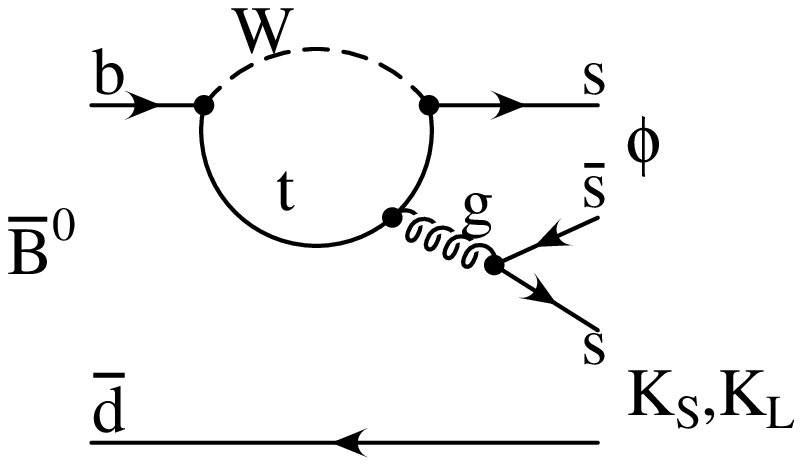}
\caption{$b\to ss\bar s$ process.} 
\label{fig:b2sssfeyn}
\end{figure}
\begin{figure}[tbp]
\centering
\includegraphics[height=50mm]{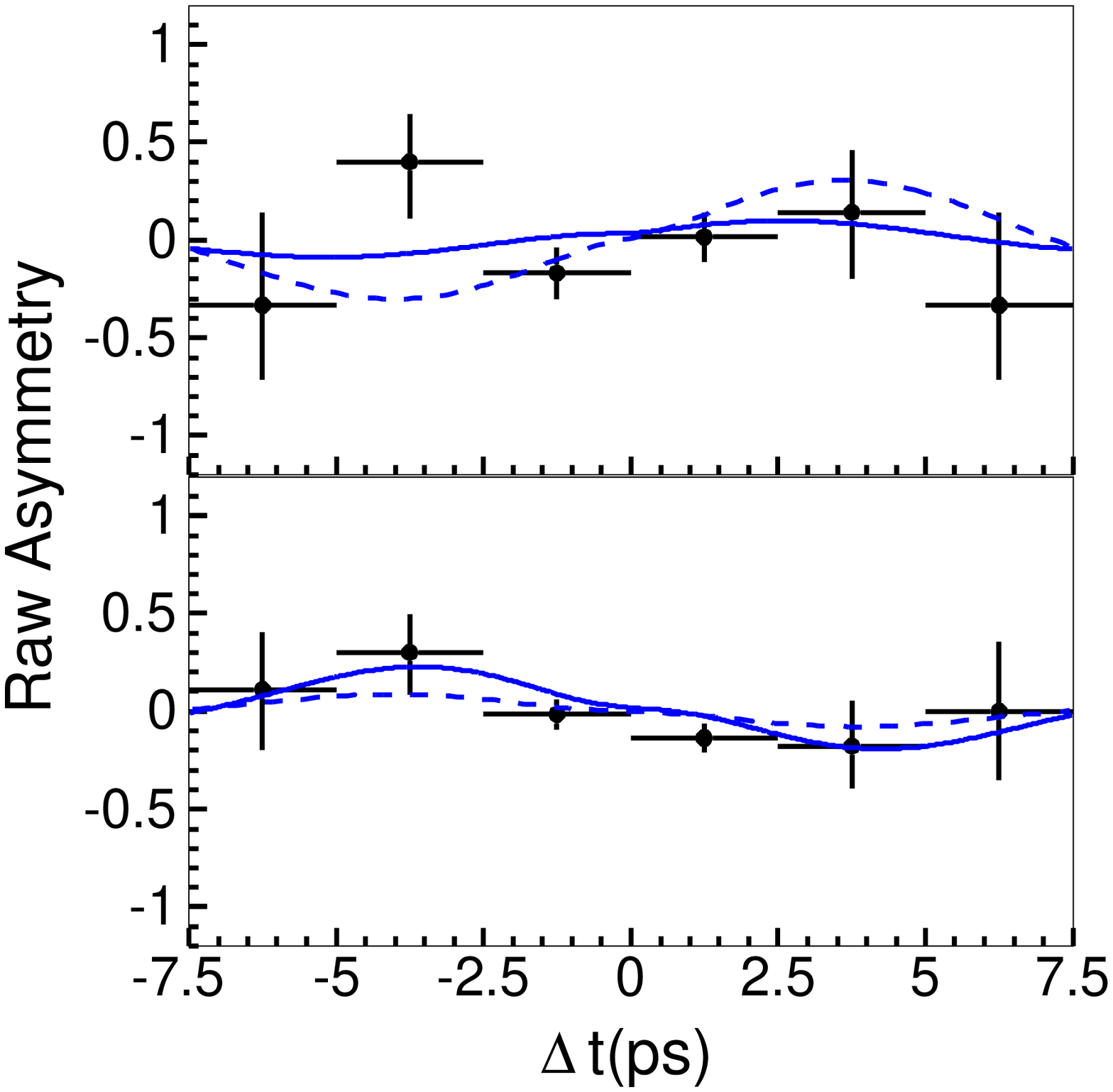}
\includegraphics[height=50mm]{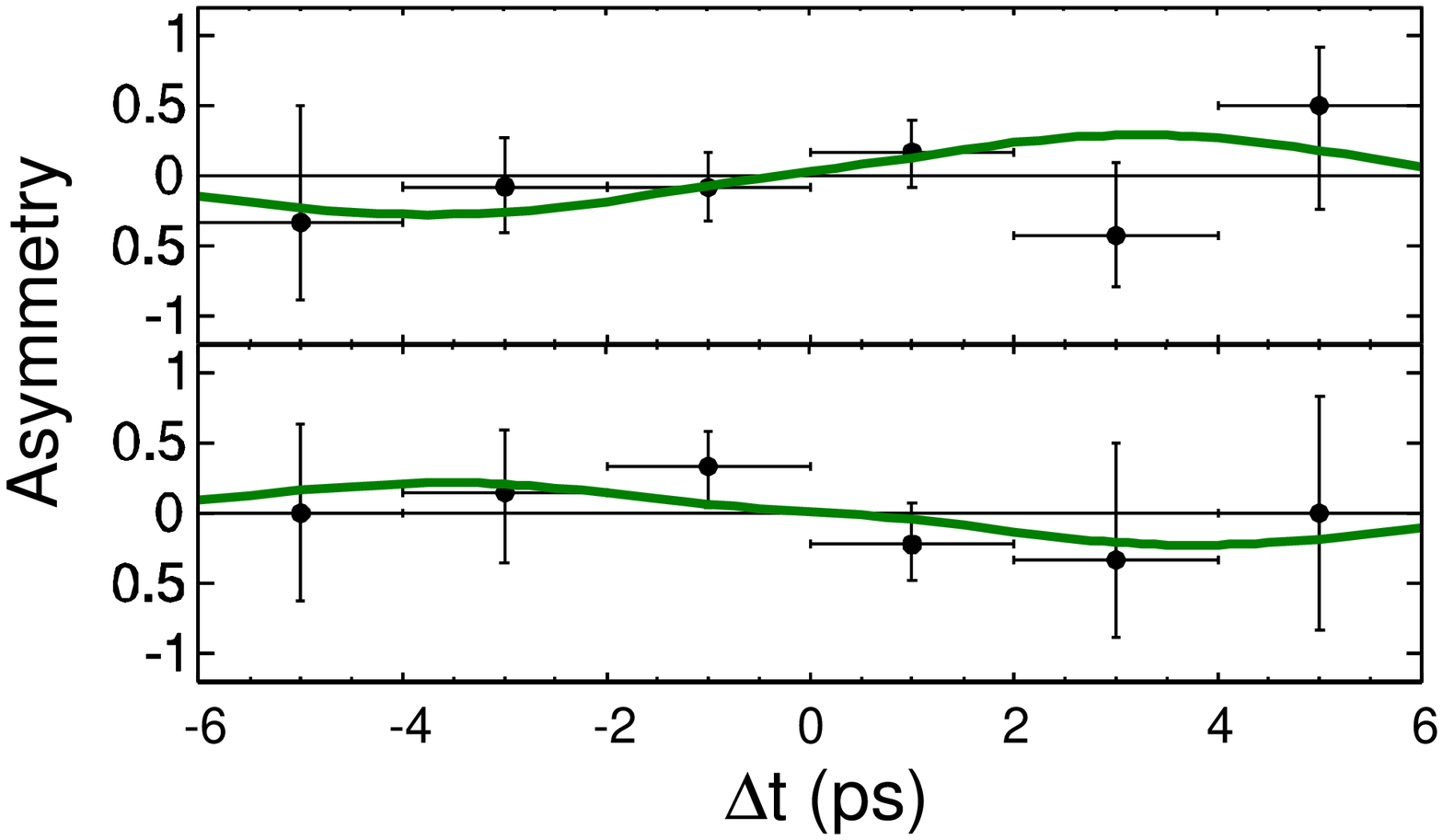}
\caption{Time-dependent CP asymmetry in decays $\bar B^0\to \phi K_{S}$(top) and $\bar B^0\to \phi K_{L}$ (bottom) for Belle (left) and Babar (right) data.
The solid lines show fitting results, and the dashed lines (Belle) give the SM expectation.} 
\label{fig:b2sss_bfac}
\end{figure}

Decays dominated by the penguin process $\bar b\to ss\bar s$ (Figure~\ref{fig:b2sssfeyn}) constitute in the SM an alternative way to measure $\sin 2\phi_1$.
In this process the three amplitudes, where the virtual quark is a $u$, $c$, or $t$, would cancel due to the unitarity of the CKM matrix were it not for the high mass of the $t$-quark.  
The amplitude involving $t$ thus dominates and is proportional to $V_{ts}V_{tb}^*\sim -A\lambda^2$, which has a real value.
By the same reasoning as with $B\to J/\psi K_S$, the time-dependent decay rate thus depends on $\sin 2\phi_1$ as Eq.~(\ref{eqn:cpasym}).
If $\bar b\to ss\bar s$ happens to include a NP process with a relative complex phase, this would be manifested as a shift in $\phi_1$.
Because the SM rate is suppressed through cancellation, any NP processes could appear at a high rate relative to the SM in this mode.
The decay $\bar B^0\to \phi K_{S,L}$ proceeds predominantly via $\bar b\to ss\bar s$, and the asymmetries are evident in the data displayed in Figure~\ref{fig:b2sss_bfac} from Belle\cite{bsss_belle} and Babar\cite{bsss_babar}.

\subsection{$b\to sq\bar q$}
\begin{figure}[tbp]
\centering
\includegraphics[height=25mm]{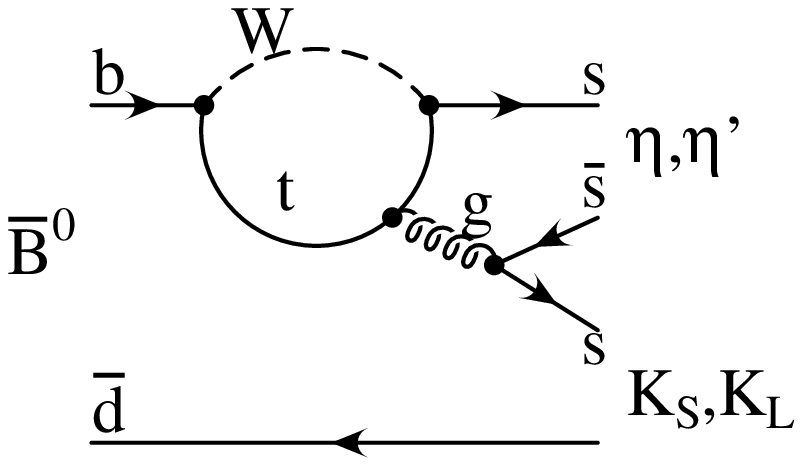}
\includegraphics[height=25mm]{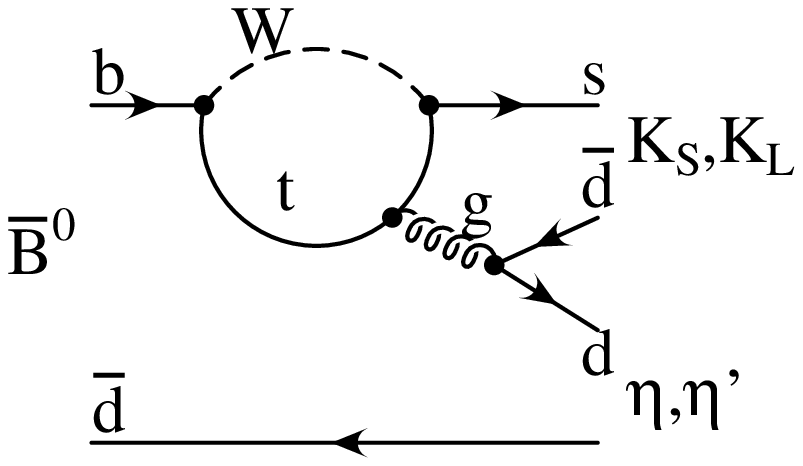}
\includegraphics[height=25mm]{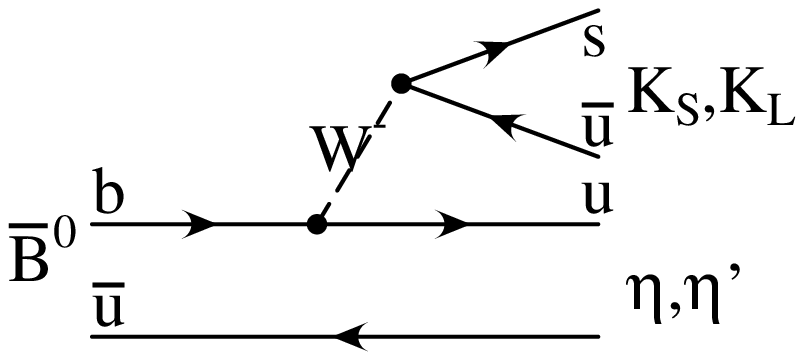}
\caption{$b\to sq\bar q$ processes.} 
\label{fig:b2sqqfeyn}
\end{figure}

\begin{figure}[tbp]
\centering
\includegraphics[height=90mm]{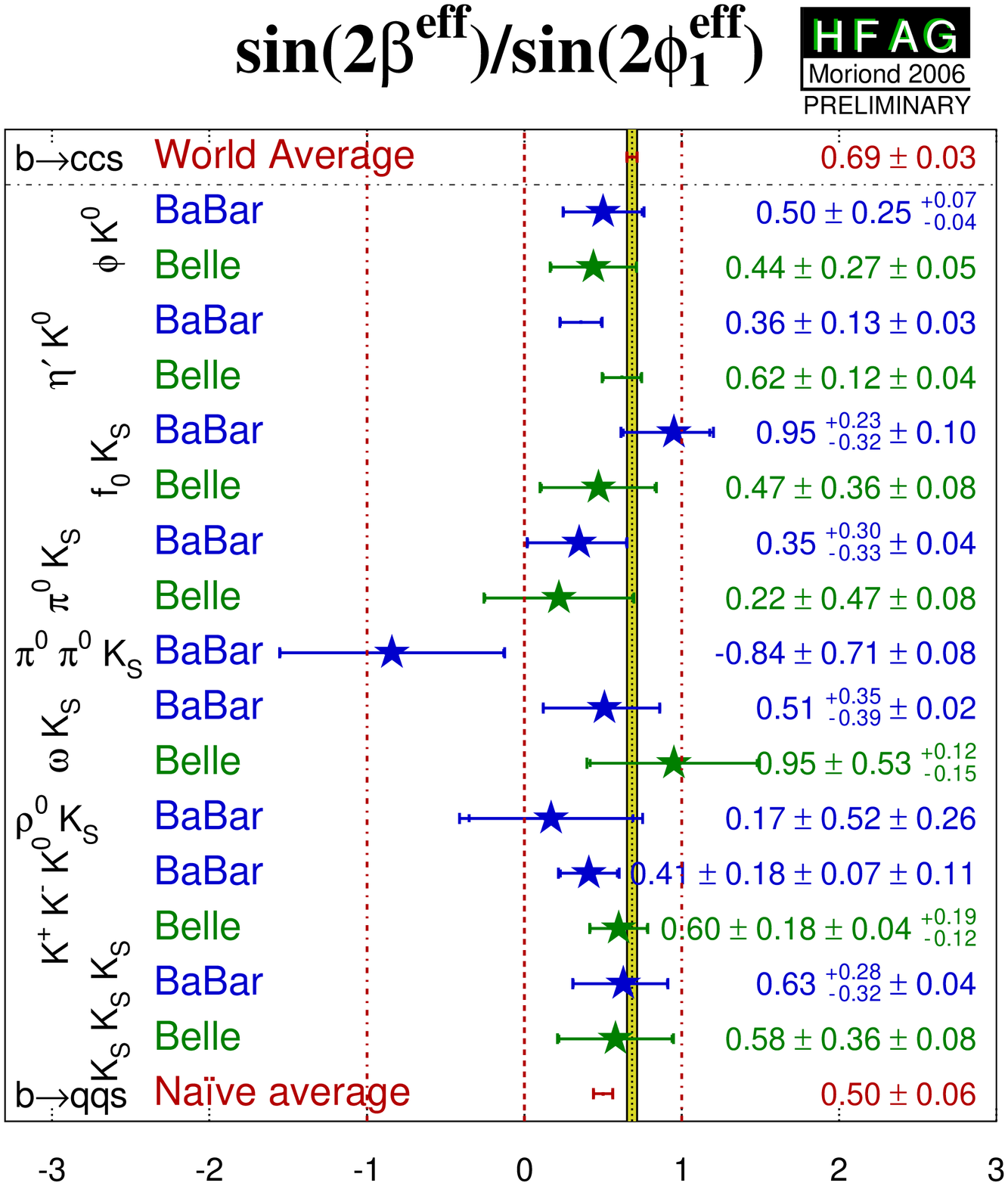}
\caption{From HFAG\cite{hfag}:  ``na\"{i}ve'' world average of $\sin 2\phi_1$ based on $b\to sq\bar q$ decays.} 
\label{fig:hfag}
\end{figure}
The decay $\bar B^0\to \eta^{(')} K_0$ may also proceed via $\bar b\to ss\bar s$. 
However, because $\eta^{(')}$ is not a pure $s\bar s$ state, $b\to sq\bar q$ processes (Figure~\ref{fig:b2sqqfeyn}) may contribute.  
Of these, the $b\to u$ channel (Figure~\ref{fig:b2sqqfeyn}(c)) involves a complex coupling and an amplitude relative to the  penguin processes that is not well known.
If the penguin amplitude is not as small as expected, the amplitude of the $CP$-asymmetry could differ from $\sin 2\phi_1$ without contributions from New Physics.

The Heavy Flavor Averaging Group (HFAG) has compiled a ``na\"{i}ve'' world average value of $\sin 2\phi_1$ based on $b\to sq\bar q$ decays and assuming that the tree contributions may be ignored.
As of this meeting the most recent HFAG value, $\sin 2\phi_1=0.50\pm 0.06$\cite{hfag}, is based on results reported at the Winter 2006 conferences (Figure~\ref{fig:hfag});
only one result has been updated ($\rho^0 K_S$\cite{babar_rho0KS}), and its change is not sufficient to affect the average.
A comparison with the average value measured in charmonium modes ($0.685\pm\pm0.032$) yields marginal agreement, with a confidence level of $9.2\times 10^{-3}$ (2.6~$\sigma$).
Because of the complication with unknown tree contributions, it is difficult to establish agreement with expectations, but the understanding is evolving, and this set of modes will continue to be watched over the next few years.

\begin{figure}[tbp]
\centering
\includegraphics[height=80mm]{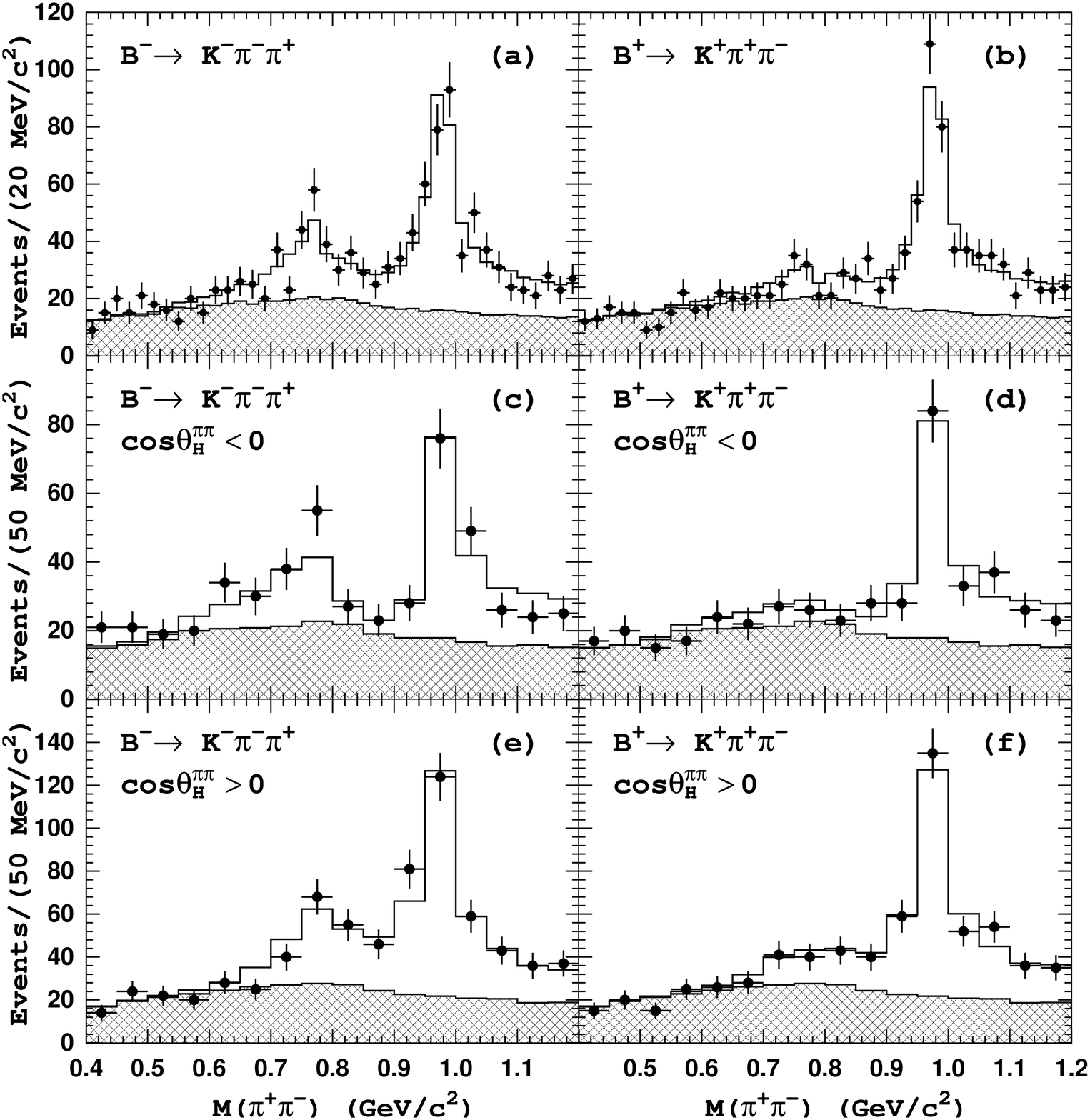}
\caption{$\pi^+\pi^-$ mass distributions for $B\to K\pi\pi$ candidates, $B^-$ (left) and $B^+$ (right), with no cuts on $\pi\pi$ helicity angle $\theta_H^{\pi\pi}$ (top), $\cos\theta_H^{\pi\pi}<0$ (center), and $\cos\theta_H^{\pi\pi}>0$ (bottom).
Data are displayed as symbols with errors, the fit is an open histogram, and the background is crosshatched.
} 
\label{fig:b2krho}
\end{figure}
\subsection{B$\to K^-\rho^0$}
Belle reports the observation of a direct CP asymmetry in the decay $B^-\to K^-\rho^0$ in a Dalitz analysis of the three-body decay  $B^-\to K^-\pi^-\pi^+$\cite{b2krho}.  
The integrated asymmetry 
$A_{CP}={N^--N^+\over N^-+N^+}$ is found to be $+0.30\pm 0.11\pm 0.02^{+0.11}_{-0.04}$ where the errors are due to statistics, experimental systematics, and decay model uncertainty, respectively.
Figure~\ref{fig:b2krho} displays the $\pi^+\pi^-$ invariant mass of $B^-\to K^-\pi^-\pi^+$ candidates, separated by charge, showing clearly the difference in rates to $\rho^0(770)$.
This is the first observation of a nonzero $CP$ asymmetry in a charged $B$ meson.
The result is generally consistent with theoretical predictions\cite{Krho-theory}.

\subsection{$B\to s\ell^+\ell^-$ and effective Wilson coefficients}
\begin{figure}[tbp]
\centering
\includegraphics[height=30mm]{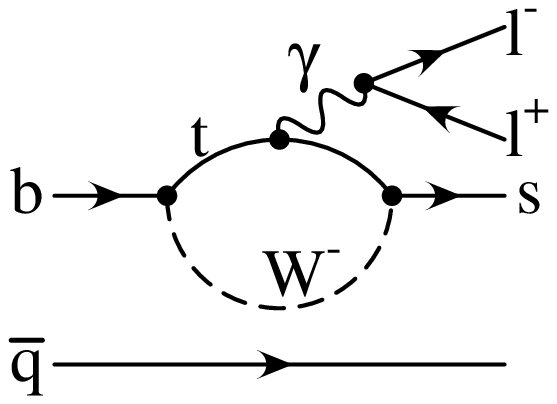}
\includegraphics[height=30mm]{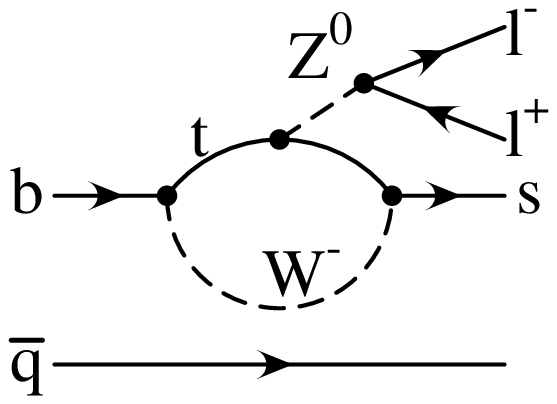}
\includegraphics[height=25mm]{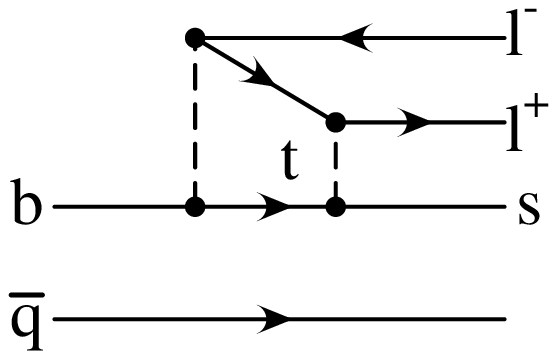}
\caption{Dominant processes in $B\to s\ell^+\ell^-$ decay.
} 
\label{fig:B2sll}
\end{figure}
The three processes which dominate decays of the type  $B\to s\ell^+\ell^-$ are shown in Figure~\ref{fig:B2sll}.
Each includes ``short-distance'' interactions, which can be treated as pointlike with perturbative corrections in powers of $\alpha_s$, and ``long-distance'' interactions requiring nonperturbative approaches. 
The Operator Product Expansion (OPE) formally separates the two regimes, enabling well-defined treatments in the calculation of theoretical amplitudes.  
Factoring out the fundamental weak couplings, the amplitude is then the product of an ``effective Wilson coefficient'' $\tilde C_i^{eff}$, characterizing the short-distance part, and a long-distance term.
The coefficients associated with the three processes of Figure~\ref{fig:B2sll} are
$\tilde C_7^{eff}$, $\tilde C_9^{eff}$, and $\tilde C_{10}^{eff}$, respectively.
They have been calculated theoretically to next-to-next-to-leading-order (NNLO) in $\alpha_s$ in the SM\cite{wilson_nnlo}.  

Previous measurements of ${\cal B}(B\to X_s\gamma)$ have constrained $\tilde C_7^{eff}$, and much of the region ($\tilde C_9^{eff},\tilde C_{10}^{eff}$) is limited by total branching fractions of $b\to s\ell^+\ell^-$ decays.  
To obtain relative signs $(\pm)$ and more direct measurements of these amplitudes, one can exploit the characteristic distributions in $q^2$, the helicity angle $\theta$, and direct $CP$ asymmetry of the three  processes.
By fitting the observed distribution one can sort out the contribution of each to obtain these three $\tilde C_i^{eff}$.
Any NP that might contribute to this mode is not only unlikely to be suppressed in the standard way, it is also unlikely to simulate the three different components in their rates and/or distributions, so one would expect to see deviations in relative signs and amplitudes of $\tilde C_i^{eff}$.

\begin{figure}[tbp]
\centering
\includegraphics[height=60mm]{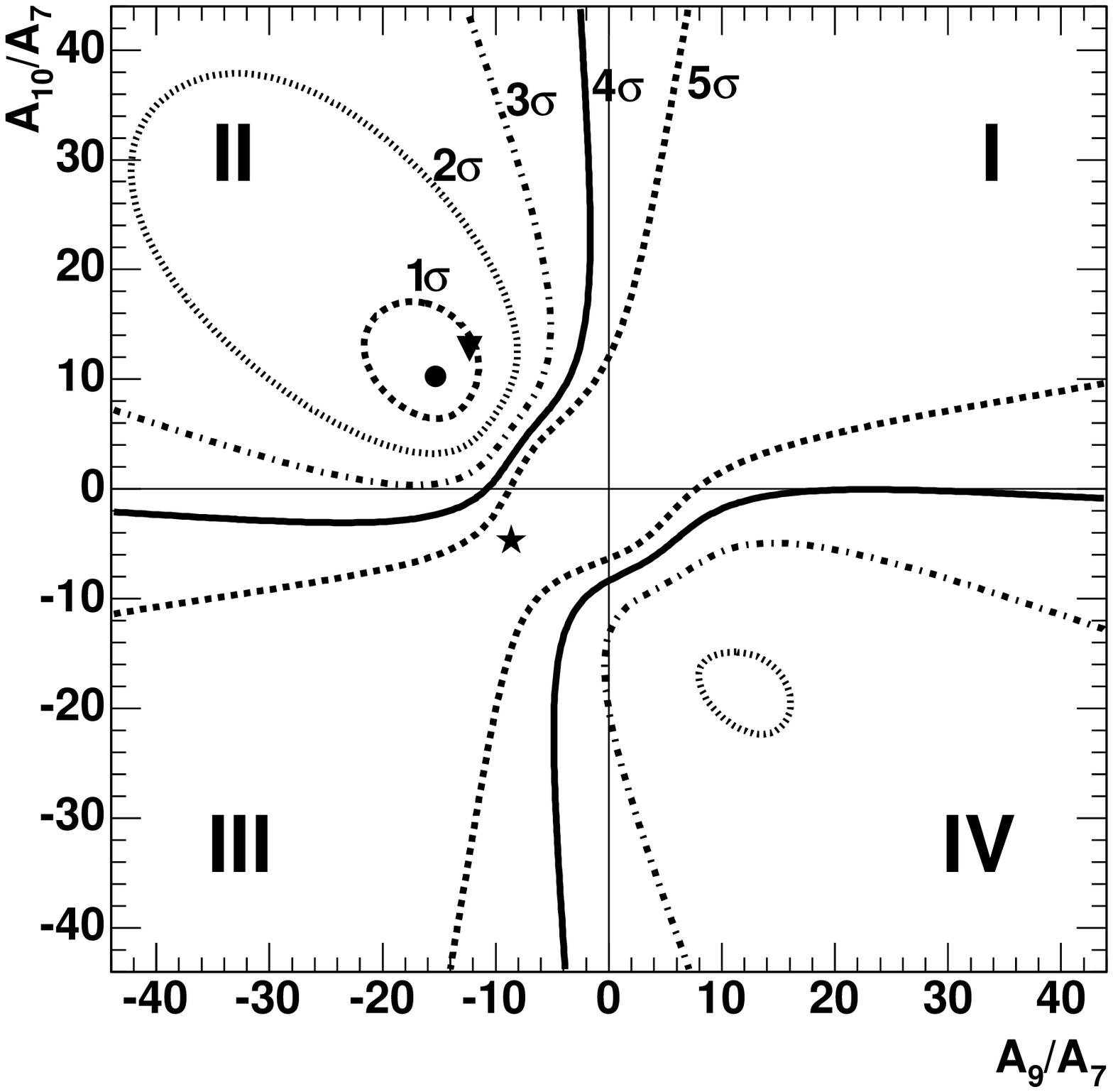}
\caption{Confidence contours for fit of $B\to K^{*}\ell^+\ell^-$ sample (Belle) to $A_9/A_7$ and $A_{10}/A_7$, where $A_7$ is taken to be negative.  Symbols indicate the result of the fit (solid circle), SM expectation (triangle), and $A_{10}$ positive (star).
} 
\label{fig:wilsoncoeff}
\end{figure}
\begin{figure}[tbp]
\centering
\includegraphics[height=50mm]{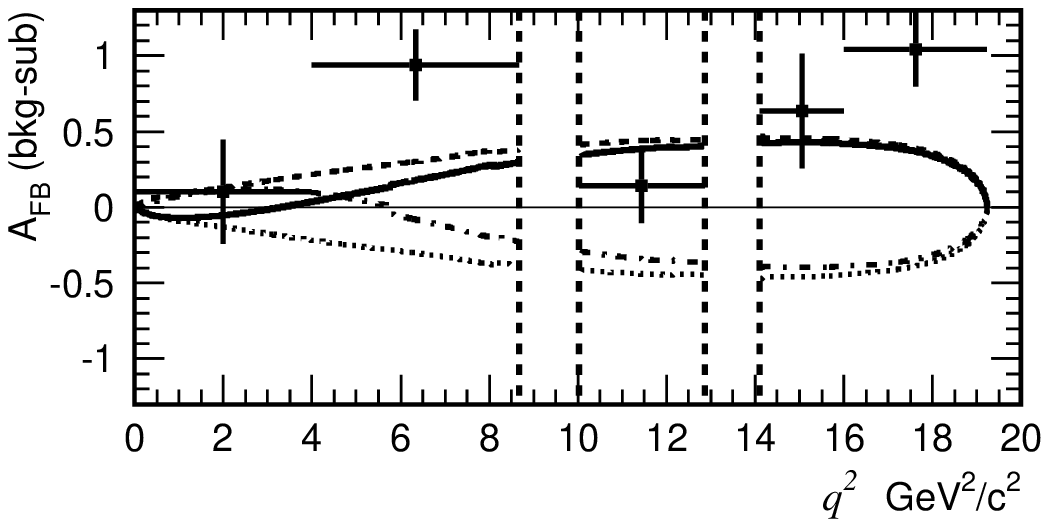}
\includegraphics[height=50mm]{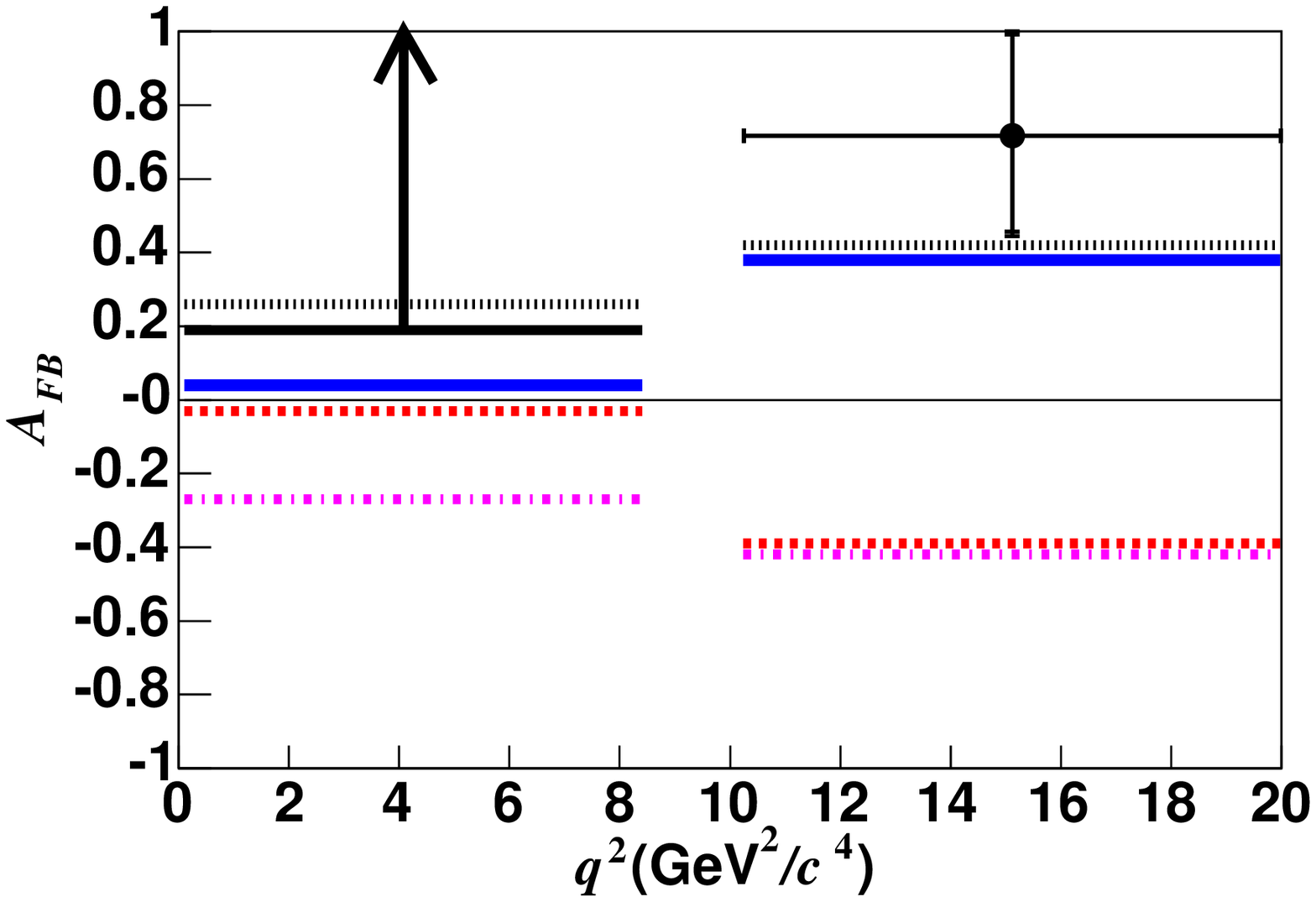}
\caption{Background-subtracted forward-backward asymmetry of $B\to K^{*}\ell^+\ell^-$ candidates at Belle (left) and Babar (right).
Curves on Belle plot are fit result for negative $A_7$ (solid), $(A_7=0.330,\ A_9=4.069,\ A_{10}=-4.213)$ (dashed), $(A_7=-.280,\ A_9=2.419,\ A_{10}=1.317)$ (dot-dashed), $(A_7=0.280,\ A_9=2.219,\ A_{10}=3.817)$ (dotted). 
On Babar plot, the SM expectation is solid blue. 
} 
\label{fig:wilson_afb}
\end{figure}
Both Belle and Babar have studied the decays $B\to K^{(*)}\ell^+\ell^-$.
The effective Wilson coefficients may be expressed as a sum of leading and higher order terms:
$\tilde C_i^{eff}=A_i+(higher\ order)_i$.
For the experimental measurement, the parameters $A_9$ and $A_{10}$ are determined through a fit to the distribution in $(q^2,\cos\theta)$, where the higher order terms fixed to the theoretical values;
because $|A_7|$ is already well constrained, its absolute value is fixed. 
Belle has looked at $3.86\times 10^8$ $B\bar B$ events and fitted the $(q^2,\cos\theta)$ distribution of $114\pm13$ $B\to K^{*}\ell^+\ell^-$ candidates to obtain
\begin{eqnarray*}
{A_9\over A_7} = -15.3^{+3.4}_{-4.8}\pm 1.1\\
{A_{10}\over A_7} = -10.3^{+5.2}_{-3.5}\pm 1.8\\
A_9\cdot A_{10}<0\  (98.2\%\ CL)
\end{eqnarray*}
The confidence contours in the plane (${A_9\over A_7},{A_{10}\over A_7}$) are shown in Figure~\ref{fig:wilsoncoeff}.
Given limited statistics, one can express the $\cos\theta$ distribution as a simple forward-backward angular asymmetry, which in the case of $B\to K^{*}\ell^+\ell^-$ is expected to vary substantially with $q^2$.
Figure~\ref{fig:wilson_afb}(left) shows this asymmetry for several bins of $q^2$ in $B\to K^{*}\ell^+\ell^-$ candidates at Belle, with curves corresponding to standard and non-standard scenarios.
A study of the same decay by Babar in $2.29\times 10^8$ $B\bar B$ events\cite{KllAFB} gives similar consistency with the SM (Figure~\ref{fig:wilson_afb}(right)).

\subsection{$b\to d\gamma$}
\begin{figure}[tbp]
\centering
\includegraphics[height=30mm]{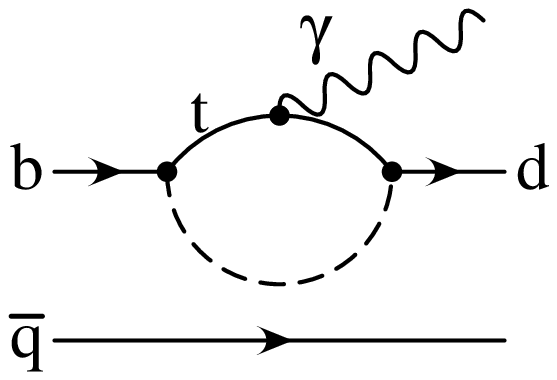}
\caption{SM process for $b\to s(d)\gamma$.} 
\label{fig:b2dgam}
\end{figure}
Decays of the type $b\to d\gamma$ are useful because the ratio of its rate to that of $b\to s\gamma$ is well-defined in the SM.
The dominant process is shown in Figure~\ref{fig:b2dgam}.  
The amplitude of $b\to d\gamma$ has three penguin contributions, where the internal quark is $u,\ c,\ {\rm or}\ t$.
If the three quark masses are equal, the unitarity of the CKM matrix results in zero amplitude due to cancellation, so the finite value is due to the large mass of the $t$-quark.
Because the decay is dominated by a single diagram that is identical to that for $b\to s\gamma$ except for one quark replacement,  their ratio (with minor corrections) is simply the ratio of the absolute squared CKM matrix elements, 
\begin{eqnarray*}
{\Gamma(b\to d\gamma)\over \Gamma(b\to s\gamma)}\sim \left| {V_{td}\over V_{ts}} \right|^2.
\end{eqnarray*}
Because hadronic uncertainties tend to cancel in the ratio, the theory errors are relatively small, of order 10\%.
As with the other loop decays, this mode is suppressed by approximate CKM cancellation and is thus sensitive to NP.

\begin{figure}[tbp]
\includegraphics[width=0.5\textwidth]{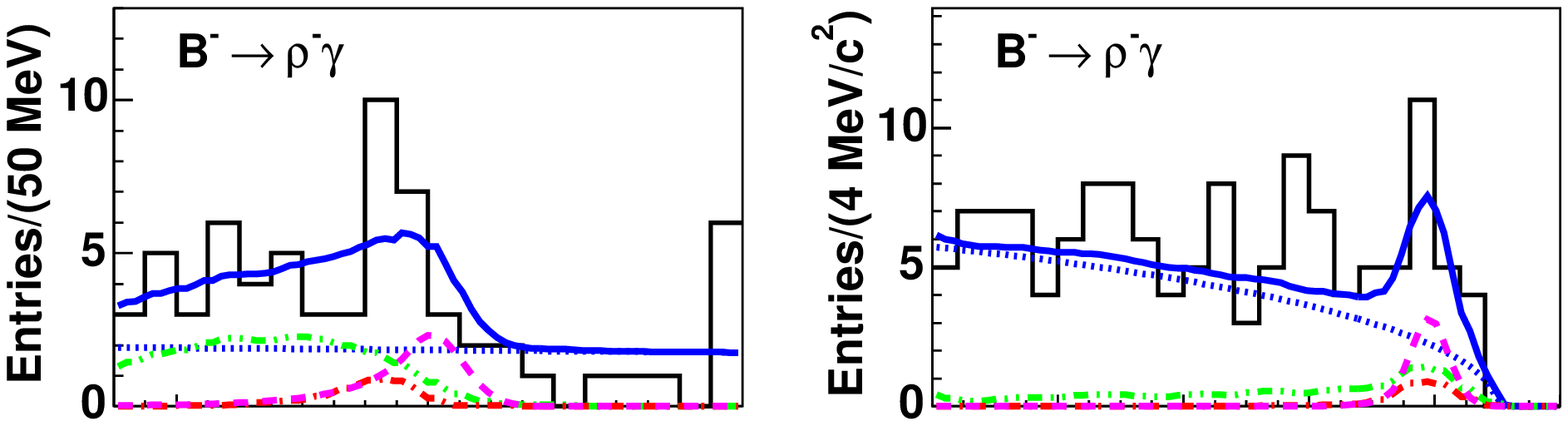}\\
\includegraphics[width=0.5\textwidth]{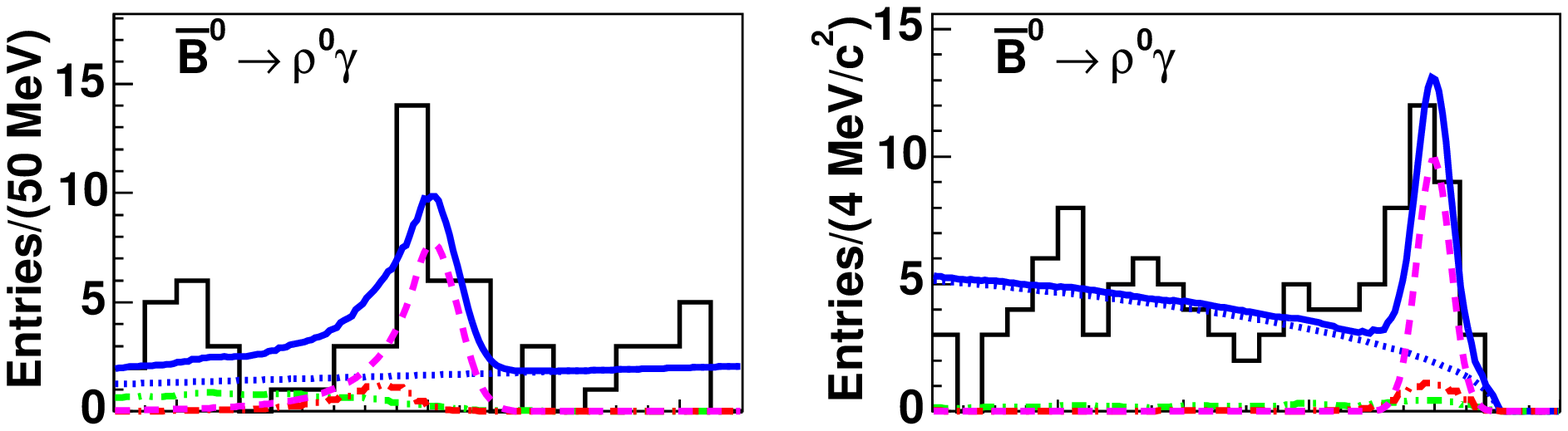}\\
\includegraphics[width=0.5\textwidth]{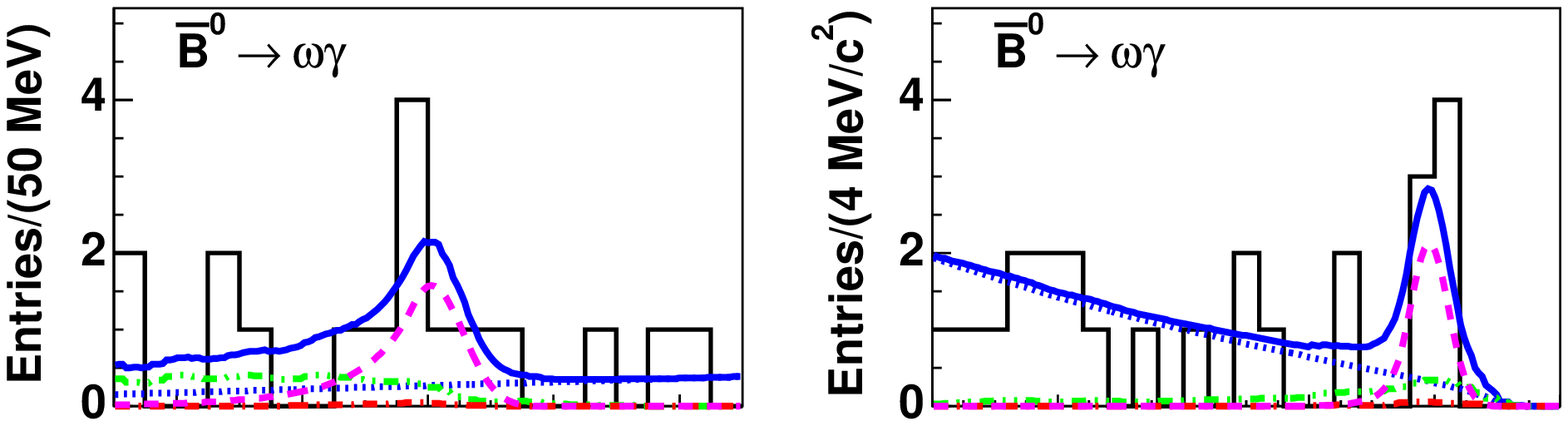}
\includegraphics[width=0.5\textwidth]{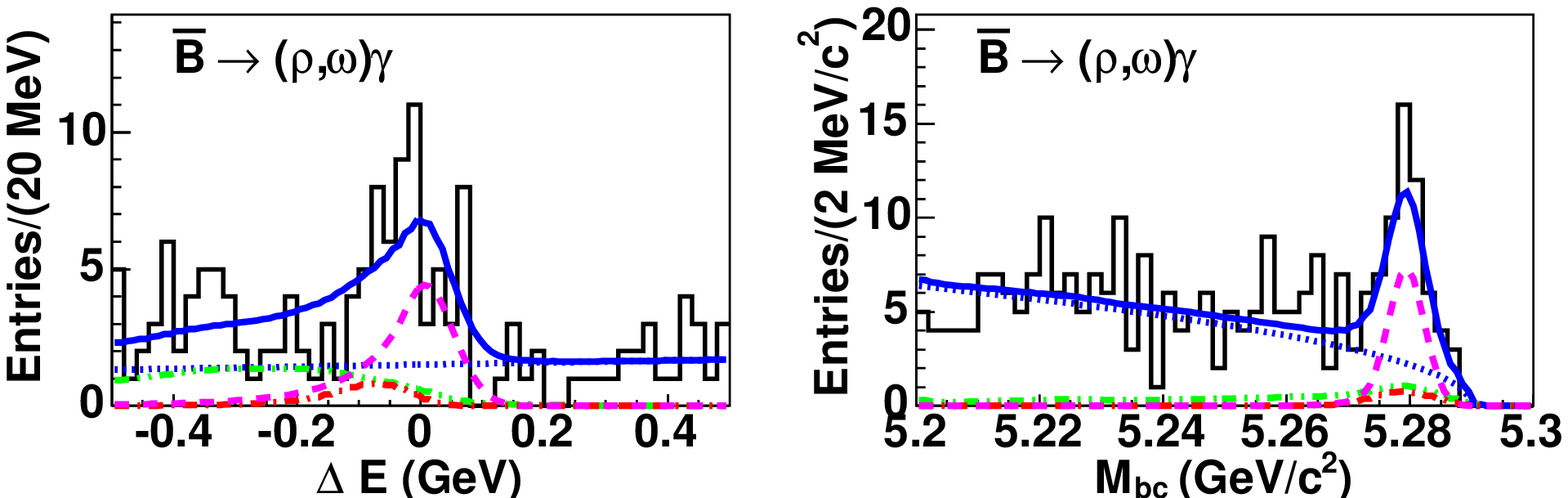}
  \caption{Candidates for $\bar B^0\to\rho^0\gamma$, $B^-\to\rho^-\gamma$, $\bar B^0\to\omega\gamma$, and combined, in 386 million $B\bar B$ events from Belle.}
  \label{fig:b2drad_data}
\end{figure}

Both Belle and Babar have searched for the exclusive modes $B\to\rho\gamma$ and $B\to\omega\gamma$.  
In 386M $B\bar B$ events, Belle has found evidence for $b\to d\gamma$ in the modes $\bar B^0\to\rho^0\gamma$, $B^-\to\rho^-\gamma$, and $\bar B^0\to\omega\gamma $  \cite{belle_b2d}.
The modes are identified through full reconstruction of 
decays in  $e^+e^-\to\Upsilon$(4S)$\to B\bar B$ events.  
Figure {\ref{fig:b2drad_data}} displays candidate distributions in $\Delta E$ and $M_{\rm bc}$.
Under an assumption of isospin invariance, 
${\cal B}(B^-\to\rho^-\gamma)=2{\tau_{B^+}\over\tau_{B^0}}{\cal B}(\bar B^0\to\rho^0\gamma)=2{\tau_{B^+}\over\tau_{B^0}}{\cal B}(B^0\to\omega\gamma)$, the three modes are combined into a single measurement of ${\cal B}(B^-\to\rho^-\gamma)$ (designated as ${\cal B}(B\to (\rho/\omega)\gamma)$) at 5.1$\sigma$ significance:
\begin{eqnarray*}
{\cal B}(B\to (\rho/\omega)\gamma)&=&(1.32{^{+0.34}_{-0.31}}{^{+0.10}_{-0.09}})\times 10^{-6}
\end{eqnarray*}
The ratio with   the corresponding $b\to s\gamma$ value,
\begin{eqnarray*}
{{\cal B}(B^-\to (\rho/\omega)\gamma)\over {\cal B}(B^-\to K^{*-}\gamma)}&=&0.032{\pm 0.008}{\pm 0.002},\\
{\rm yields}\ \ \ \ \ \ \ \ \ \ \ \ \ \ \ \ \ \ \ \ 
\left|{V_{td}\over V_{ts}}\right|&=&0.199{^{+0.026}_{-0.025}}{^{+0.018}_{-0.015}}.
\end{eqnarray*}
Babar's search for the same modes among 211M~$B\bar B$ events\cite{babar_b2d} yields possible signals at the 2.1$\sigma$ level and the following 90\% confidence upper limits:
\begin{eqnarray*}
{\cal B}(B^-\to (\rho/\omega)\gamma)<1.2\times 10^{-6},\ 
{{\cal B}(B^-\to (\rho/\omega)\gamma)\over {\cal B}(B^-\to K^{*-}\gamma)}<0.029,\ 
\left|{V_{td}\over V_{ts}}\right|<0.19.
\end{eqnarray*}

\subsection{ $B^+\to \tau^+\nu$}
\begin{figure}[t]
\centering
\includegraphics[width=50mm]{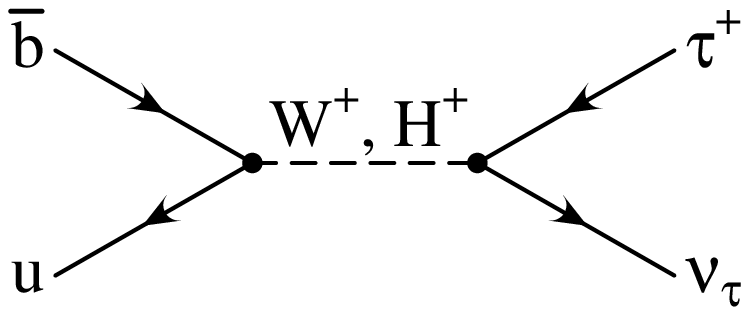}
\includegraphics[width=40mm]{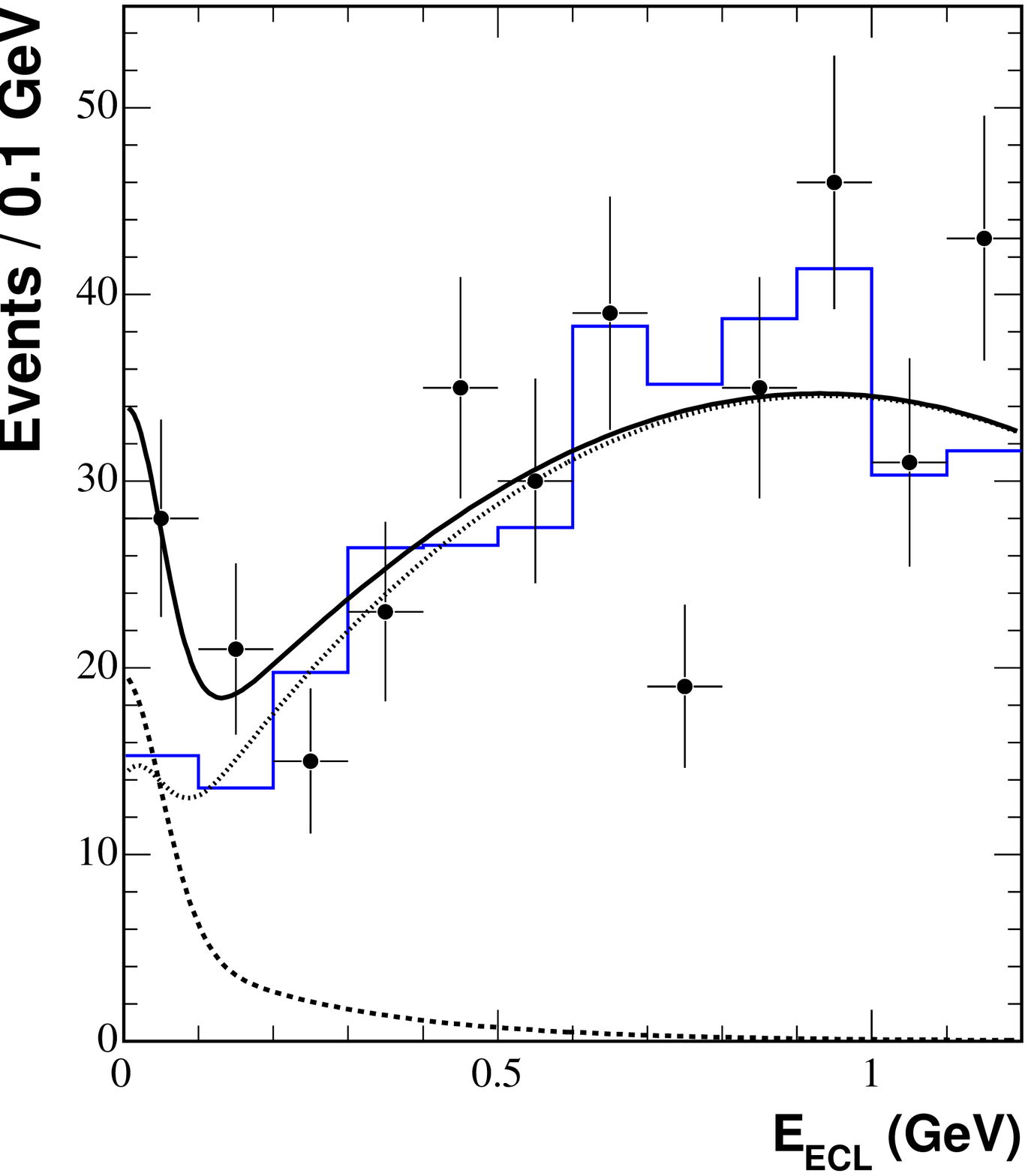}
\includegraphics[width=45mm]{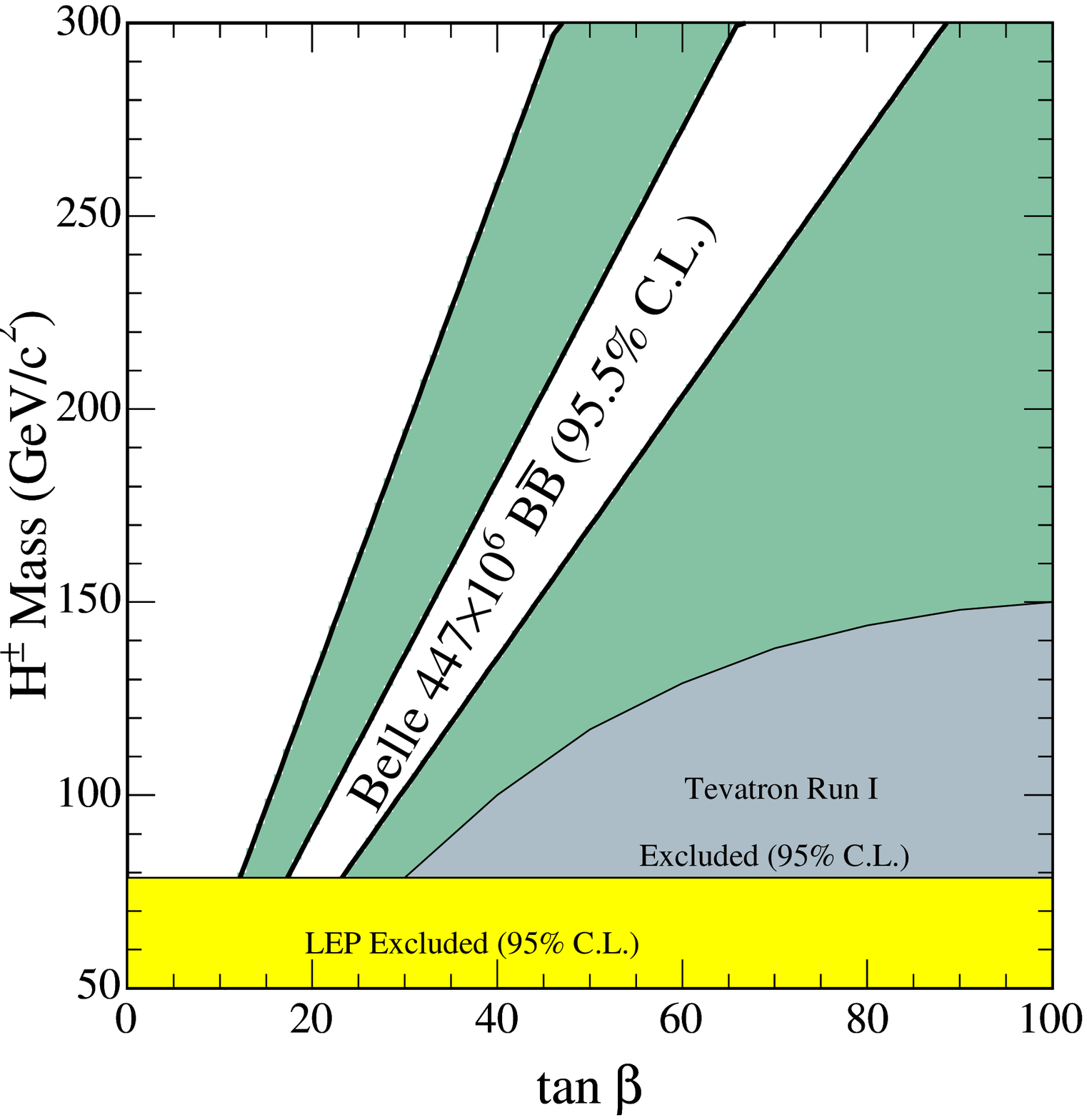}
\caption{(left) SM process for the decay $B^+\to \tau^+\nu$. (center) The distribution of residual calorimeter energy in candidate events at Belle.
(right) Region of charged Higgs mass still allowed after Belle measurement of ${\cal B}(B^+\to \tau^+\nu_\tau)$ (white).
} 
\label{fig:taunu_E}
\end{figure}

From a theoretical point of view, this mode proceeds through a clean process (Figure~\ref{fig:taunu_E}(left))  that provides a measure of $|V_{ub}|$ with relatively small uncertainties and is independent of measurements made in spectator (tree) decays:
\begin{eqnarray*}
{\cal B}(B^+\to \tau^+\nu_\tau)={G_F^2m_B\over 8\pi}m_\tau^2\left(1-{m_\tau^2\over m_B^2}\right)^2f_B^2|V_{ub}|^2\tau_B.
\end{eqnarray*}
Based on the current knowledge of $|V_{ub}|$, the branching fraction is predicted to be $(1.59\pm 0.40)\times 10^{-4}$.
Experimentally, this mode is nontrivial to reconstruct, as it always involves at least two neutrinos.
Belle and Babar have conducted searches by first fully reconstructing a $B^+$ meson and then searching in the rest of the event for $\tau$ decay products with no residual tracks or energy.
Belle has searched in $447\times 10^6$ $B\bar B$ events for $\tau^-\to \mu^-\bar\nu_\mu\nu_\tau$, $e^-\bar\nu_e\nu_\tau$, $\pi^-\nu_\tau$, $\pi^-\pi^0\nu_\tau$, and $\pi^-\pi^+\pi^-\nu_\tau$ on the signal side, which constitute 81\% of all tau channels.
Figure~\ref{fig:taunu_E}(center) shows the distribution of residual calorimeter energy in candidate events.
There is an excess at zero which is interpreted as first evidence for $B^+\to \tau^+\nu$, with a branching fraction
${\cal B}(B^+\to \tau^+\nu_\tau)=(1.06{^{+0.34}_{-0.28}}{^{+0.18}_{-0.16}})\times 10^{-4}$\cite{belle-taunu}.
Babar has also conducted a search, in $232\times 10^6$ $B\bar B$ events, and obtained an upper limit at 90\% confidence of $2.6\times 10^{-4}$\cite{babar-taunu}.

Assuming that there is no New Physics influence on ${\cal B}(B^+\to \tau^+\nu_\tau)$, one can use the measured value to extract $|V_{ub}|$.  
The CKMFitter group has performed this calculation and has combined it with measured values of $\Delta m_d$ to constrain $(\rho,\eta)$, comparing with constraints based on CP asymmetries\cite{ckmfitter}.

Any New Physics that may appear in this channel is limited by the observed consistency between the theoretical and experimental values.
The ratio of the branching fraction reported by Belle and the theoretical value would differ from unity as\cite{hou_higgs}
\begin{eqnarray*}
r_H=(1-{m_B^2\over m_H^2}\tan^2\beta)^2
\end{eqnarray*}
in the two-Higgs doublet model, where $r_H\equiv{{\cal B}(B^+\to \tau^+\nu_\tau)\over  {\cal B}(B^+\to \tau^+\nu_\tau)_{SM}}$ and $\tan\beta$ is the ratio of vacuum expectation values of the two doublets.
Figure~\ref{fig:taunu_E}(right) shows in white the ($\tan\beta,M_H)$ region not excluded by the Belle measurement.

\subsection{$B^0\to \tau^+ \tau^-$}

\begin{figure}[t]
\centering
\includegraphics[height=35mm]{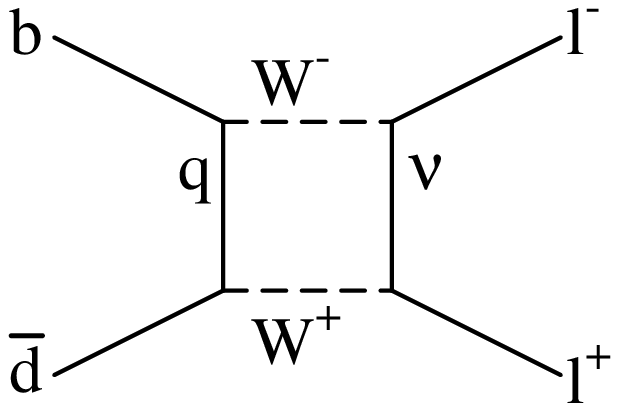}
\includegraphics[height=35mm]{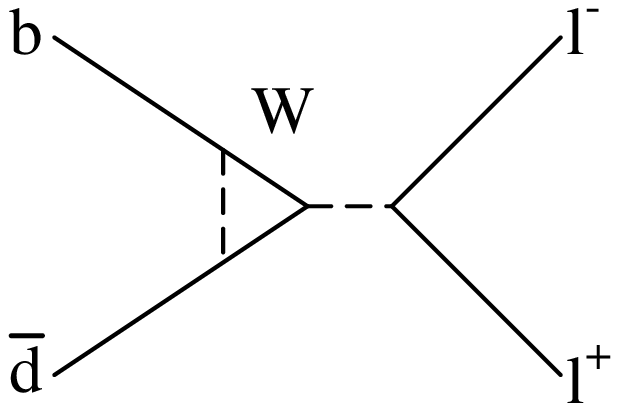}
\caption{SM processes for $B^0\to \tau^+ \tau^-$.} 
\label{fig:b2tautau}
\end{figure}
The SM prediction for ${\cal B}(B^0\to \tau^+ \tau^-)$ is $\sim 2\times 10^{-7}$\cite{SM_b2tautau}, proceeding via the processes shown in Figure~\ref{fig:b2tautau}.
Again, due to CKM suppression of the SM expectation, this mode is sensitive to BSM models, in this case ones that involve direct lepton-quark couplings.

Babar has searched for this mode in  $232\times 10^6$ $B\bar B$ events\cite{babar_tautau}.
Events in which one hadronic $B^0$ decay is fully reconstructed are examined for evidence of tau pairs in the modes $\tau\to \mu\nu_\mu\nu_\tau,\  e\nu_\mu\nu_\tau,\ \pi\nu_\tau,$ and $\rho\nu_\tau$.
The analysis includes requirements that track charges be consistent with the hypothesis, that there be no extra tracks and that the residual calorimeter energy be consistent with zero.
A limit at 90\% confidence level of ${\cal B}(B^0\to \tau^+ \tau^-)<4.1\times 10^{-3}$ is obtained.

\subsection{$B_d$, $B_s$ $\to\gamma\gamma$}
\begin{figure}[t]
\centering
\includegraphics[height=30mm]{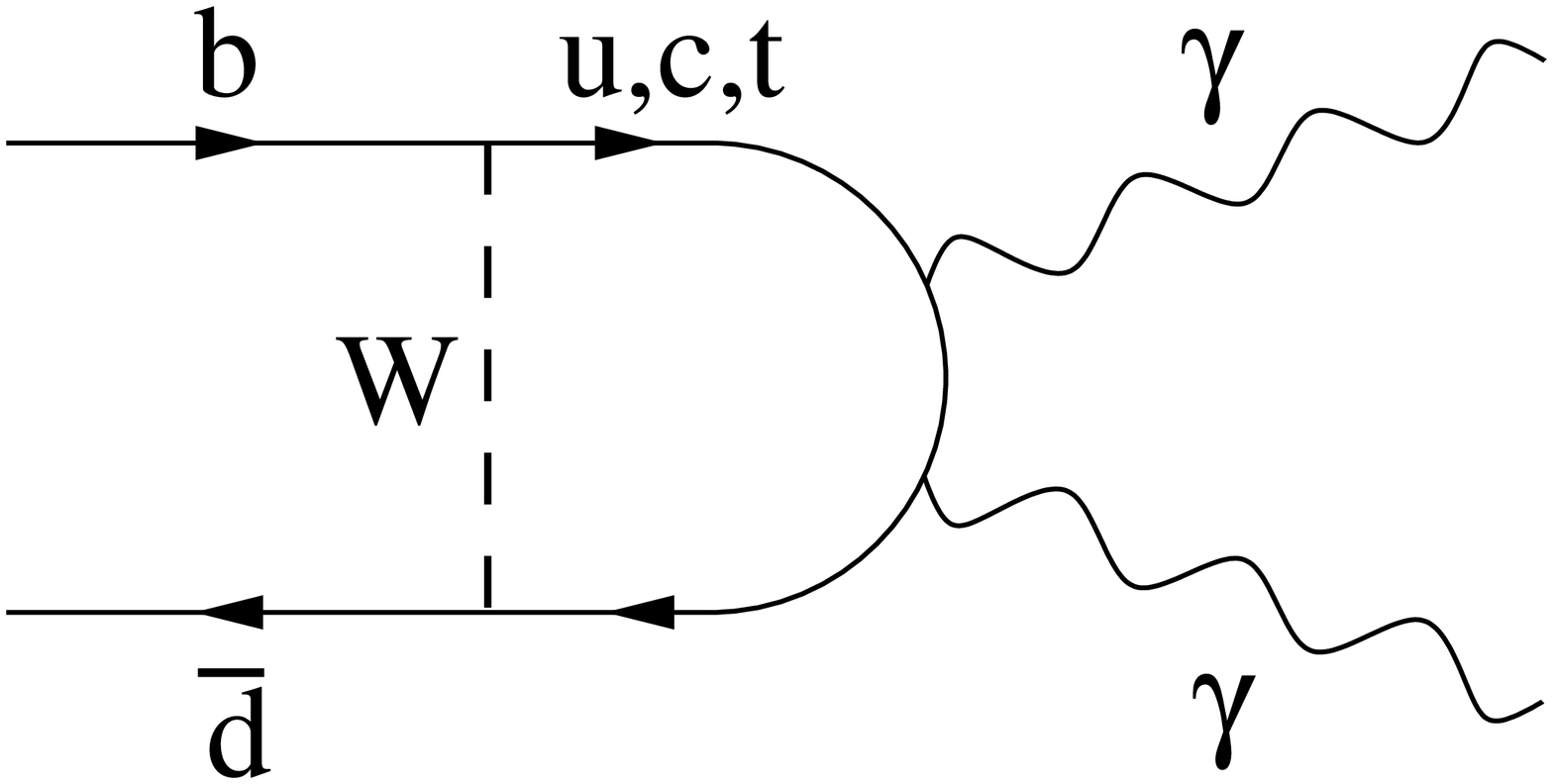}
\caption{SM process for $B_d\to \gamma\gamma$.} 
\label{fig:b2gamgam}
\end{figure}
The decays $B_{d(s)} \to\gamma\gamma$ may occur through the loop annihilation process shown in Figure~\ref{fig:b2gamgam}.
SM branching fractions are estimated at ${\cal B}(B_d\to\gamma\gamma)\sim 3\times 10^{-8}$ and ${\cal B}(B_s\to\gamma\gamma)\sim 0.5-1.0\times 10^{-6}$\cite{b2gamgam_theory}.
In some BSM models, this rate may be enhanced by as much as two orders of magnitude\cite{b2gamgam_bsm}.

A search by Belle in $111\times 10^6$ $B\bar B$ events collected at the $\Upsilon$(4S) resonance yields a 90\% CL upper limit ${\cal B}(B_{d} \to\gamma\gamma)<6.2\times 10^{-7}$\cite{belle_b2gamgam}.
Belle has also searched in $9.0\times 10^4$ $B_s^{(*)}\bar B_s^{(*)}$ events collected at the $\Upsilon$(5S) resonance and finds ${\cal B}(B_{s} \to\gamma\gamma)<5.6\times 10^{-5}(90\%\ CL)$\cite{belle_bs2gamgam}.
This value is three times smaller than the previous best limit.

\subsection{Charm and tau summary}
Although much of the attention at a $B$-factory is directed at $B$ decays, significant constraints on NP can be gained from charm and tau decays.  
In the charm sector the rates of mixing and flavor-changing neutral currents (FCNC), and CP violation are very small in the SM due to CKM cancellations, providing additional opportunities to observe NP.  
Tau leptons are well understood in the SM and their decays may be examined for  violation of flavor, lepton number, and/or baryon number.
In particular the events are rather clean, so high sensitivity can be achieved for neutrinoless decays which can be fully reconstructed.

Listed below are a selection of such results reported this year.

\begin{itemize}	
\item Search for FCNC (Babar, 263~fb$^{-1}$) in decays of $D^+$, $D_s^+$, and $\Lambda_c^+$ to $h^+\ell^+\ell^-$ where $h^+$ is a charged hadron.  The search covered 20 modes, 17 of which resulted in new limits\cite{babar_fcnc}.

\item Search for mixing in decays $D^0\to K^+\pi^-$ (Belle, 400~fb$^{-1}$), where the flavor of the $D$ is tagged by reconstruction of $D^{*+}\to D^0\pi^+$.
By taking the decay time distribution, mixing is separated from doubly-Cabibbo-suppressed decays (DCSD).
A new limit on the integrated mixing rate, $R_M< 4\times 10^{-4}$ (95\% CL), has been set\cite{belle_RM}.
The expected rate based on the SM is $\sim 10^{-4}$.

\item Search for mixing based on $D^0\to K^+\pi^-\pi^0$ (Babar, 230~fb$^{-1}$), where the flavor of the $D$ is tagged by reconstruction of $D^{*+}\to D^0\pi^+$.  A Dalitz analysis of the decay time distribution provides improved separation between mixing and DCSD.
The limit obtained is $R_M< 5.4\times 10^{-4}$ (95\% CL) \cite{babar_RM}.

\item New limits on radiative $\tau$ decays: ${\cal B}(\tau\to e\gamma)<1.1\times 10^{-7}$ (90\% CL) (Babar\cite{babar_egam}), ${\cal B}(\tau\to e\gamma)<1.2\times 10^{-7}$ (90\% CL) (Belle\cite{belle_egam}), ${\cal B}(\tau\to \mu\gamma)<4.5\times 10^{-8}$ (90\% CL) (Belle\cite{belle_egam}).
Out of these comes a new constraint on the minimal supersymmetric model, which predicts
${\cal B}(\tau\to\mu\gamma)=3.0\times 10^{-6}\times({\tan\beta\over 60})^2({M_{SUSY}\over 1\ TeV})^{-4}$.

\item Limits on baryonic decays of $\tau$: ${\cal B}(\tau\to \bar\Lambda\pi^+)<1.4\times 10^{-7}$ (90\% CL) and ${\cal B}(\tau\to \Lambda\pi^-)<0.72\times 10^{-7}$ (90\% CL) (Belle\cite{belle_Lampi}).

\item Limits on $\tau\to\ell h^+h^-$, $\ell V^0$: where $h^\pm-$ comprises $\pi^\pm$, $K^\pm$ and $V^0$ are $\rho^0$, $K^*(892)^0$, or $\phi$ decays to $h^+h^-$.  The 90\% CL limits on the various branching fractions are in the range $(1-8)\times 10^{-7}$ (Belle\cite{belle_lhh}).

\item Searches for $\tau\to\ell K_S$: ${\cal B}(\tau\to e K_S)<5.6\times 10^{-8}$, ${\cal B}(\tau\to \mu K_S)<4.9\times 10^{-8}$ (90\% CL)  (Belle\cite{belle_eKS})

\end{itemize}

\section{SUMMARY}

The $B$-factories, which were designed to make precision measurements of CKM parameters, are now able to take advantage of the tremendous volume of accumulated data to search for New Physics.
NP may appear in many different guises, as disagreements between CKM parameters measured through different processes or in the appearance of forbidden or suppressed decays.
Because many hadronic weak decays are suppressed beyond their nominal coupling strengths due to the unitarity of the CKM matrix, which results in the approximate cancellation of amplitudes, the sensitivity to NP in many cases is comparable and complementary to that at higher energy machines. 
I have presented here a selection of recent results based on $B$, charm, and tau decays from Belle and Babar.

\begin{acknowledgments}
The author wishes to thank the SSI organizers and staff for their hospitality.
This work is supported by Department of Energy grant \# DE-FG02-84ER40153.
\end{acknowledgments}


\end{document}